\documentclass[twocolumn,aps,pra,reprint,superscriptaddress,nofootinbib]{revtex4-1}

\usepackage{amsmath}
\usepackage{bm}
\usepackage{mathrsfs}
\usepackage{hyperref}
\hypersetup{citecolor=blue}
\usepackage{color}
\usepackage{graphicx}
\usepackage{subfigure}
\usepackage{verbatim}
\begin{document}

\preprint{PRA - v0.4a}

\title{A unitary quantum lattice algorithm for two dimensional quantum turbulence}

\author{Bo Zhang}
\author{George Vahala}

\affiliation{Department of Physics, William \& Mary, Williamsburg, VA 23185}
\author{Linda Vahala}
\affiliation{Department of Electrical \& Computer Engineering, Old Dominion University, Norfolk, VA 23529}
\author{Min Soe}
\affiliation{Department of Mathematics  and Physical Sciences, Rogers State University, Claremore, OK 74017}


\date{\today}

\begin{abstract}
Quantum vortex structures and energy cascades are examined for two dimensional quantum turbulence (2D QT) at zero temperature. A special unitary evolution algorithm, the quantum lattice algorithm (QLA),  is employed to simulate the Bose-Einstein condensate (BEC) governed by the Gross-Pitaevskii (GP) equation. A parameter regime is uncovered in which, as in 3D QT, there is a short Poincar\'e recurrence time.  It is demonstrated that such short recurrence times are destroyed by stronger nonlinear interaction. The similar loss of Poincar\'e recurrence is also seen in the 3D GP equation \cite{yepezeuro}. 
Various initial conditions are considered in an attempt to discern if 2D QT exhibits inverse cascades are is seen in 2D CT (classical turbulence).
In our simulation parameter regimes, no dual cascades spectra were observed for 2D QT--- unlike that seen in 2D  CT.
\end{abstract}


\maketitle

\section{Introduction}\label{sec:intro}
A thorough understanding of vortex motion in turbulence remains an important challenge \cite{feynmantb}. Now in classical turbulence, the vortex can not be precisely defined and is viewed as a continuous property of the fluid.  However in quantum turbulence the quantum vortex is a topological singularity with discrete quantized circulation. 
The size of the quantized (isolated) vortex is characterized by the coherent length in superfluid, 
a measure of the relative strength of the  kinetic energy.
 This makes quantum turbulence, loosely defined as the dynamics of tangled vortices in 3D or of regularly/irregularly distributed vortex points in 2D \cite{feynmantb}, a suitable prototype for understanding classical turbulence \cite{tsubota2009intro,vinen2006}, particularly in the very long wavelength limit in which the discrete quantization of the vortices should become unimportant. 
Additionally, since the turbulence driven by  a BEC gas expansion displays different characteristics  
from those of a trapped BEC gas \cite{henn09}, there is much still to be understood in the dynamics of QT. 

The evolution of  quantum turbulence in a 2D BEC gas at zero temperature is well described by the mean field GP equation for the one particle wave function $\psi$:
\begin{equation}\label{eq:gpe}
i\partial_t\psi=-\nabla^2\psi+g|\psi|^2\psi,
\end{equation}
where $g$ is the nonlinear coupling constant for the weak interactions among the ground state bosons, in lattice units where $\hbar=1$ and $m=1/2$. Here we omit the chemical potential term $-\mu\psi$ because this term only adds a global phase shift $e^{-i\mu t}$ to the wave function.  
One important feature of the GP system is that it is a Hamiltonian system with exact conservation of energy.

  If the phase space dynamics of an energy conserving system is bounded then there will be a Poincar\'e recurrence of the initial state. For continuous systems, the Poincar\'e recurrence period $T_P$ is typically so extremely long that for all practical purposes it can be almost considered to be infinite. However, for certain discrete maps like the Arnold's cat map, $T_P$ can be unexpectedly very short. We shall find that there is a class of initial conditions for the 2D GP equation for which the Poincare recurrence time is on the order of $\mathcal{O}[10^4]$ iterations on a $512^2$ grid.  This Poincare recurrence time is independent of the dimensionality of the dynamics - whether 2D or 3D \cite{yepezeuro}.  Moreover the Poincare recurrence time scales with diffusion ordering:  as the (linear) grid is increased from $L_1$ to $L_2$, the Poincare recurrence time $T_P$ increases by a factor of $(L_2/L_1)^2$ - again independent of the dimensionality of the dynamics. 
For a short Poincar\'e recurrence, the interaction is required to be much smaller than the kinetic energy. Under this condition, the BEC gas is highly dilute and the recurrence time can be very short.

\subsection{Superfluid theory}

The simplest theory for a BEC in the zero-temperature limit is given by Lagrangian density for a scalar complex field $\psi$:
\begin{equation}
\label{eq:gplag}
\mathscr{L}=\frac{i}{2}\left(\psi^{\dagger}\partial_t\psi-\psi\partial_t\psi^{\dagger}\right)-\nabla\psi^{\dagger}\cdot\nabla\psi-\frac{g}{2}\left(\psi^{\dagger}\psi\right)^2.
\end{equation}
%
%
The invariance of $\mathscr{L}$ with respect to phase rotation, to space  and to time translations yields the conservation of density, momentum and energy, respectively.
  The resulting set of hydrodynamic equations are:
%
\begin{subequations}
\begin{eqnarray}
\partial_t\rho+\nabla\cdot(\rho\bm{v})&=&0
\label{eq:continuity}\\
\rho(\partial_t\bm{v}+\bm{v}\cdot\nabla\bm{v})&=&-2\rho\nabla\left(g\,\rho-\frac{\nabla^2\sqrt{\rho}}{\sqrt{\rho}}\right),
\label{eq:eulerfluid}
\end{eqnarray}
\end{subequations}
where $\rho$ is the probability density and $\rho\bm{v}$ the probability momentum:
\begin{subequations}
\label{eq:hydrodef}
\begin{eqnarray}
\rho&\equiv&\psi^{\dagger}\psi\\
\rho\bm{v}&\equiv& i(\psi\nabla\psi^{\dagger}-\psi^{\dagger}\nabla\psi).
\end{eqnarray}
\end{subequations}
Equation (\ref{eq:continuity}) is identified with the continuity equation of fluid; while (\ref{eq:eulerfluid}) is recognized as the Euler equation for a barotropic inviscid fluid. Notice that the pressure now consists of both local and non-local terms: the normal (local) pressure $g\rho^2$, and the quantum pressure $\displaystyle -\frac{\nabla^2\sqrt{\rho}}{\sqrt{\rho}}$ which is non-local. To analyze the energy spectra of quantum turbulence, the (constant) total energy
\begin{equation}\label{eq:tote}
E_T=\int d{x}^2 \left( \frac{1}{2}\rho|\bm{v}|^2+g\rho^2+2|\nabla\sqrt{\rho}|^2 \right)
\end{equation}
(in 2 spatial dimensions) is decomposed into three components:
%
\begin{subequations}
\label{eq:engdef}
\begin{eqnarray}
\text{kinetic energy:}&\quad& E_K=\frac{1}{2}\int d{x}^2\,\rho|\bm{v}|^2\\
\text{internal energy:}&\quad& E_I=g\,\int d{x}^2\,\rho^2\\
\text{quantum energy:}&\quad& E_Q=2 \int d{x}^2\,|\nabla\sqrt{\rho}|^2.
\end{eqnarray}
\end{subequations}
One important feature of quantum turbulence is its compressibility. Much of the classical turbulence literature has focussed on incompressible turbulence but there is  sound wave emission \cite{leadbeater2001} in a BEC gas. Therefore, to compare quantum turbulence with its classical counterpart, we further decompose the  kinetic energy into its compressible and incompressible components. This is achieved by introducing the density weighted velocity $\mathbf{q}=\sqrt{\rho}\,\bm{v}$ \cite{nore1997,tsubota2010} and the Helmholtz decomposition of a non-singular vector field. The longitudinal component of $\mathbf{q}$ contributes to the compressible kinetic energy while the transverse component contributes to the incompressible kinetic energy:
%
\begin{subequations}
\label{eq:inccomp}
\begin{eqnarray}
\text{compressible:}&\quad E_C=(2\pi)^2\int d{k}^2\,|\tilde{\mathbf{q}}_c|^2;\\
\text{incompressible:}&\quad E_{IC}=(2\pi)^2\int d{k}^2\, |\tilde{\mathbf{q}}_{ic}|^2;
\end{eqnarray}
\end{subequations}
where $\tilde{\mathbf{q}}_c$ and $\tilde{\mathbf{q}}_{ic}$ are defined as:
%
\begin{subequations}
\label{eq:kinedecomp}
\begin{eqnarray}
\tilde{\mathbf{q}}_c&\equiv& \frac{\mathbf{k}\cdot\tilde{\mathbf{q}}}{k^2}\mathbf{k};\\
\tilde{\mathbf{q}}_{ic}&\equiv& \tilde{\mathbf{q}}-\frac{\mathbf{k}\cdot\tilde{\mathbf{q}}}{k^2}\mathbf{k};
\end{eqnarray}
\end{subequations}
with $\tilde{\mathbf{q}} = \tilde{\mathbf{q}}_c + \tilde{\mathbf{q}}_{ic} $ being the Fourier transform of $\mathbf{q}$. Consequently, the energy density of incompressible kinetic energy and compressible kinetic energy can be defined as 
\begin{equation}\label{eq:spectra}
\varepsilon_{c,ic}(k)=k\int_0^{2\pi} d\theta\,|\tilde{\mathbf{q}}_{c,ic}(k,\theta)|^2,
\end{equation}
using polar coordinates $k, \theta$.

For 2D classical incompressible turbulence, the conservation (in the inviscid limit) of both the enstrophy,  $Z=\int d\mathbf{r}\,|\nabla\times\bm{v}|^2$, and energy has a profound effect on energy transfer. Based on the assumption of incompressibility, isotropy and self-similarity, Kraichnan \cite{kraichnan1967} and Batchelor \cite{batchelor1969} have demonstrated that there exists dual cascades in the inertial range, with an inverse cascade of kinetic energy but a direct cascade of enstrophy in wave number space with spectral exponents for the kinetic energy spectrum:
\begin{align*}
\text{inverse cascade:}&\quad \varepsilon_{ic}(k)\propto k^{-5/3};\\
\text{direct cascade:}&\quad \varepsilon_{ic}(k)\propto k^{-3}.
\end{align*}
The existence of dual cascades is a hallmark of 2D (classical) turbulence.  In 3D classical turbulence, enstrophy (in the inviscid limit) is no longer conserved and one only finds the direct cascade of kinetic energy leading to the well known Kolmogorov $k^{-5/3}$ spectrum.  In 2D, the inverse cascade dictates that the energy is transferred to larger spatial scales, which is manifested by the coalescence of eddies (with the same rotational sense) into larger and larger vortices.  On the other hand, the direct cascade determines that the vorticity is cascaded to smaller scales and dissipated away through viscosity at the end of the inertial limit.  However, for quantum turbulence, vortices can be created via nucleation
 \cite{takeuchi10} and annihilated through the mutual interaction leading to the fusion of $\emph{counter-rotating}$ pairs, presumably due to the quantum pressure in Eq. (3b).  Thus enstrophy in 2D QT is not conserved. Lacking the conservation of enstrophy and because of compressibility effects,  2D quantum turbulence does not necessitate the existence of dual cascades. 
  
  The recent papers on 2D quantum turbulence by Horng et al. \cite{horng2009} and Numasato et al. \cite{tsubota2010}, have examined the energy cascades.  Horng et. a.\cite{horng2009} consider the initial conditions of a Gaussian BEC cloud with embedded vortices in an external strong potential well.  In their quasi-stationary turbulent state they see a clear demarcation between the spatial distribution of the compressible kinetic energy from that of the incompressible energy.  In particular, the quantum vortices are predominantly located outside the Thomas Fermi radius along with the incompressible kinetic energy distribution, while the compressible kinetic energy distribution is predominantly located within the Thomas Fermi radius. 
 They find a dual cascade of the incompressible kinetic energy spectrum---akin to 2D classical turbulence:  an inverse energy cascade with spectrum $k^{\alpha}$  with $\alpha \sim -5/3$ (but with large error bars in a very restricted k-range) and a direct enstrophy cascade with (a noisy) spectral exponent $\alpha \sim -4$.  This deviates from the standard 2D classical direct enstrophy cascade exponent $\alpha = -3$.  Numasato \cite{tsubota2010}, on the other hand, have no external trapping potential and the initial state is spatially homogeneous with random phase in momentum space.   In contrast to Horng et al., they find a direct cascade of incompressible kinetic energy with a transient spectral exponent $\alpha \sim -5/3$ (also with large error bars and quite a narrow time window).  In their final asymptotic state, the vortices disappear from their system with the incompressible kinetic energy tending to zero.

\subsection{Organization}

This paper is organized as follows. In Sec.~\ref{sect:QLA}, we give only a brief description of the QLA algorithm since the 3D algorithm has been given in considerable detail elsewhere \cite{yepez0902}. In Sec.~\ref{sect:poin}, we observe the Poincar\'e recurrence under two very different sets of initial conditions:  in the first case when there are initially quantum vortices present in the system, and in the second case where the initial density is uniform but the wave function has random phase fluctuations. Thus in this case, there are initially no vortices, but the high kinetic energy will rapidly lead to their creation.  By gradually increasing the ratio of the interacton energy to the kinetic energy, we demonstrate the destruction of this fast Poincar\'e recurrence. 
 In Sec.~\ref{sect:spec}, we examine the power law of the energy cascades. A $k^{-3}$ power law in the incompressible kinetic energy is observed, while the compressible kinetic energy spectrum is somewhat more complicated and, as in 3D QT , exhibits multi-cascade behavior. However, this incompressible power law is absent during the evolution of the turbulence when the vortices are annihilated from the system and it reappears when vortices are re-created. It is suggestive to correlate the behavior of the large $k$-range with this $k^{-3}$ power law spectrum to that found by Nore et al. \cite{nore1997} on taking Fourier transform of an isolated quantum vortex.  Of course, there are some differences:  for the isolated quantum vortex there is no compressible kinetic energy, while from our 2D QT simulations in the large $k$ region the compressible kinetic energy spectrum is also $k^{-3}$ and of similar magnitude to the incompressible spectrum.

A close examination of the incompressible energy spectra, in our chosen parameter regimes, does not reveal the existence of the Komolgorov $k^{-5/3}$ spectrum which characterizes the inverse energy cascade in 2D classical incompressible turbulence.  In Sec.~\ref{sect:conclu} we summarize some of our conclusions.

\section{Quantum lattice algorithm}\label{sect:QLA}
We shall find that the GP system can be recovered from a simple two qubit  QLA algorithm with basis set: $|00\rangle,|01\rangle,|10\rangle,|11\rangle$ at each grid point. Applying a series of interleaved local unitary collision and unitary streaming operations at each lattice site, the QLA algorithm models the dynamics of GP system in the long wavelength limit on taking moments. To recover the GP for the single particle wave function\cite{yepez02,yepez0902} one need only consider just two of these basis set elements. Consequently, the collision operator $\hat{\mathcal{C}}$ and streaming operator $\hat{\mathcal{S}}$ can be reduced to $2\times2$ matrices. The $\sqrt{\text{SWAP}}$ gate is chosen as the collision operator in our simulation:
\begin{equation}\label{eq:collop}
\hat{\mathcal{C}}=\frac{1}{2}\begin{pmatrix}
1-i&1+i\\
1+i&1-i
\end{pmatrix},
\end{equation}
with $\hat{\mathcal{C}}^4=I$, the identity operator.  The streaming operator is simply a shift operator defined by:
\begin{align}\label{eq:strmop}
\hat{\mathcal{S}}_{\Delta x_i,0}&=n+e^{\Delta x_i\partial_{x_i}}\bar{n};\\
\hat{\mathcal{S}}_{\Delta x_i,1}&=\bar{n}+e^{\Delta x_i\partial_{x_i}}n;
\end{align}
where $\displaystyle n=\frac{1}{2}(1-\sigma_z)$, $\displaystyle \bar{n}=\frac{1}{2}(1+\sigma_z)$. $\sigma_z$ is the standard Pauli spin matrix. $\Delta x_i$ is the displacement along the lattice $\mathbf{e}_i$ direction. The subscript $\alpha$ in $\hat{\mathcal{S}}_{\Delta x_i,\alpha}$ indicates  the particular qubit being streamed. Interleaving the non-commuting operators $\hat{\mathcal{C}}$ and $\hat{\mathcal{S}}$, one obtains the unitary evolution operator $\hat{\mathcal{U}}$ for 2D GP system \cite{vahala2005}:
\begin{align}\label{eq:unitop}
\hat{\mathcal{U}}_{\alpha}&=\hat{\mathcal{I}}^2_{x,\alpha}\hat{\mathcal{I}}^2_{y,\alpha};\\
\hat{\mathcal{I}}_{x_i,\alpha}&=\hat{\mathcal{S}}_{-\Delta x_i,\alpha}\hat{\mathcal{C}}\hat{\mathcal{S}}_{\Delta x_i,\alpha}\hat{\mathcal{C}}.
\end{align}
The GP wave function is recovered from the qubit representation by $\psi(\mathbf{r},t)=q_0(\mathbf{r},t)+q_1(\mathbf{r},t)$, where $q_{\beta}$ is the (complex) probability amplitude of $|01\rangle$ and $|10\rangle$, for $\beta = 0$ or $1$ respectively. On applying this evolution operator to this spinor state:
\begin{equation}
\begin{pmatrix}
q_0(t+\Delta t)\\
q_1(t+\Delta t)
\end{pmatrix}=\hat{\mathcal{U}}_{\alpha}\begin{pmatrix}
q_0(t)\\
q_1(t)
\end{pmatrix},
\end{equation}
one obtains the dynamical evolution of the wave function \cite{yepez0902} 
\begin{equation}\label{eq:QLAevolu1}
\psi(\mathbf{r},t+\Delta t)=\psi(\mathbf{r},t)+i\Delta x^2\frac{1}{2}\nabla^2\psi(\mathbf{r},t)+(-1)^{\alpha}\mathcal{O}[\Delta x^3].
\end{equation}
on performing a Taylor expansion with respect to $\Delta x$.
Under diffusion ordering $\Delta x^2 \sim \Delta t$, this equation recovers the free particle Sch\"ordinger equation up to order $\mathcal{O}[\Delta x^3]$.  (It should be noted that the equation itself is ostensibly of order $\mathcal{O}[\Delta x^2]$ -- so that the numerical error term is of order $\mathcal{O}[\Delta x]$.) Now the error terms take the opposite sign for the evolution operator affiliated with the different spinor component. Hence by applying both evolution operators on the spinor state $\displaystyle q\equiv\bigl( \begin{smallmatrix} q_0\\q_1\end{smallmatrix} \bigr)$: 
\begin{equation}\label{eq:u0u1op}
q(t+\Delta t)=\hat{\mathcal{U}}_1\hat{\mathcal{U}}_0\,q(t), 
\end{equation}
the error itself can be further reduced to order $\mathcal{O}[\Delta x^2]$. 

To incorporate the effect of a potential, we introduce the following unitary operator:
\begin{equation}\label{eq:potop}
\Omega[V(\mathbf{r},t)]=e^{-i\Delta t\,V(\mathbf{r},t)}.
\end{equation}
$V(\mathbf{r})$ will be defined to be the nonlinear term in GP equation: $g|\psi|^2$. With this implementation, the GP equation can be reproduced up to the order $\mathcal{O}[\Delta x^2]$ through the following unitary operation:
\begin{align}\label{eq:QLA2gpe}
\psi(t+\Delta t)&=q_0(t+\Delta t)+q_1(t+\Delta t);\nonumber\\
q(t+\Delta t)&=\hat{\mathcal{U}}_1\Omega[V(t+\Delta t/2)/2]\hat{\mathcal{U}}_0\Omega[V(t)/2] q(t),
\end{align}
where the potential $V(t+\Delta t/2)$ is computed from the $\psi(t+\Delta t/2)$ obtained via the one evolution operation $\hat{\mathcal{U}}_0\Omega[V(t)/2]$.
Note that diffusion ordering is used in obtaining GP equation from QLA. This adds one additional parameter $a\equiv\Delta t$ in the GP equation to reflect the spatial/temporal mesh of the lattice. Consequently, the GP equation simulated by Eq.(\ref{eq:QLA2gpe}):
\begin{equation}\label{eq:gplattice}
i\partial_t\psi=-\nabla^2\psi+a\,g\,|\psi|^2\psi+\mathcal{O}[a^2].
\end{equation}

Due to the unitarity of QLA algorithm, the norm of the spinor is automatically conserved. However, in the standard QLA algorithm there is a very small numerical loss in the mean density during the simulation.  This is largely due to the overlapping of the two components of the spinor:
\begin{align}
\delta\bar{\rho}=\int d\mathbf{r}(|\psi|^2-|q|^2)&=\int d\mathbf{r}\,(|q_0+q_1|^2-|q_0|^2-|q_1|^2)\nonumber\\
&=\int d\mathbf{r}(q_0^{*}q_1+q_0 q^{*}_1).
\end{align}
If $q_0q_1$ is kept purely imaginary during the simulation, the overlap between the two components vanishes and the mean density is exactly conserved. To achieve this goal, two modifications are introduced to our previous QLA algorithm:

\vspace{0.5em}
\noindent
(1) the initialization of the spinor is chosen to be $q_0=\Re[\psi]$, $q_1=i\, \Im[\psi]$;\\

\vspace{0.5em}
\noindent
(2) the potential operator $\Omega$ is replaced by the non-diagonal matrix
\begin{equation}
\label{eq:newpotop}
\Omega_N=
e^{-i \sigma_x V \Delta t}=\begin{pmatrix}
\cos[V\Delta t]&& -i\sin[V\Delta t]\\
-i \sin[V\Delta t]&& \cos[V\Delta t]
\end{pmatrix},
\end{equation}
such that $\displaystyle \sum_\gamma (\Omega_N\cdot q)_{\gamma}=e^{-i V\Delta t}\psi$ in order to reproduce GP equation. 

The QLA algorithm consists of three main components: collision, streaming and local phase rotation to introduce the potential. It is clear that streaming does not alter the phase of $q_0\,q_1$, e.g. $q_0\,q_1$ will remain purely imaginary. Since the collision operator $\hat{\mathcal{C}}$ is the $\sqrt{\text{\sc swap}}$ gate, then applying this collision operation four times results in the identity operator:
\begin{equation}
\hat{\mathcal{C}}^4\begin{pmatrix}
a\\
i b
\end{pmatrix}=\hat{\mathcal{C}}^2\begin{pmatrix}
i b\\
a
\end{pmatrix}=\begin{pmatrix}
a\\
i b
\end{pmatrix};
\end{equation}
where $a$ and $b$ are real.  In actually applying the full QLA algorithm, we note that the collide-stream unitary operators do not commute.  However, what is critical is that the streaming operator never changes the phases of the spinor components so that
\begin{equation}
\hat{\mathcal{C}} \hat{\mathcal{S}^{-1}} \hat{\mathcal{C}} \hat{\mathcal{S}} \hat{\mathcal{C}} \hat{\mathcal{S}^{-1}} \hat{\mathcal{C}} \hat{\mathcal{S}} \begin{pmatrix}
a\\
i b
\end{pmatrix}=
\begin{pmatrix}
e\\
i f
\end{pmatrix}
\end{equation}
for some real $e$ and $f$.
Finally, the modified potential operator $\Omega_N$ does not alter the phase of either component of the spinor:
\begin{equation}
\Omega_N\begin{pmatrix}
a\\
i b
\end{pmatrix}=\begin{pmatrix}
a \cos[V\Delta t] + b \sin[V\Delta t] \\
- i a \sin[V\Delta t] + i b \cos[V\Delta t]
\end{pmatrix}.
\end{equation} 
As a result, the averaged density 
\begin{equation}
|\psi|^2=|e^{-i V\Delta t}(a+i b)|^2=a^2+b^2
\end{equation} is conserved exactly.
\section{Poincar\'e Recurrence}\label{sect:poin}
To probe the Poincar\'e recurrence in the 2D GP system, we introduce the time averaged ratio of the internal energy to the kinetic energy defined in (\ref{eq:engdef}) as
\begin{equation}
\displaystyle \gamma=\left\langle \frac{E_I(t)}{E_K(t)}\right\rangle
\end{equation}
 We shall find that $\gamma$ plays a crucial role in determining the existence of short Poincar\'e recurrence times. Another useful quantity is the density weighted vorticity  \cite{horng2009}: 
 \begin{equation}
 \omega_q=(\nabla\times[\sqrt{\rho}\bm{v}])\cdot\mathbf{e}_z
\end{equation} 
with $\rho$ and $\bm{v}$ being defined in (\ref{eq:hydrodef}). The two components of $\omega_q$ are:
\begin{itemize}\setlength{\itemsep}{-0.3cm}
\item $(\nabla\sqrt{\rho})\times\bm{v}\cdot\mathbf{e}_z$, which reflects the variation of the density near isolated vortex cores or near entangled vortices (i.e., vortices that strongly interact with each other);\\
\item $\sqrt{\rho}\,\nabla\times\bm{v}\cdot\mathbf{e}_z=\sqrt{\rho}\,\Gamma\,\delta(\mathbf{r}-\mathbf{r}_0)$, which pinpoints the 2D location of the vortices with their circulation $\Gamma$.
\end{itemize}
Another convenient way to visualize the vortices will be to identifying the branch cuts and singularities in a phase plot of the wave function $\psi$. In particular, at the location of a vortex there will be a phase change of $2 \pi$ per winding number, giving the impression of a branch point and the consequent branch cut.  To reflect the variation in the total number of vortices, the density weighted enstrophy \cite{horng2009}
\begin{equation}
Z=\int d\mathbf{r}\,|\omega_q|^2 
\end{equation}
is also calculated in the simulation. 

The color scheme for the distribution plots is ``thermal": \emph{blue} stands for the low values while \emph{ red} represent the high values. 
\subsection{Vortices initially imbedded in a Gaussian BEC cloud}\label{sec:vortinit}
\begin{figure*}[!h!t!b!p]
	\subfigure[$|\psi_{tot}|$,t=0]{\includegraphics[scale=0.41]{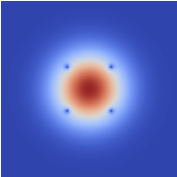}}
	\subfigure[t=8000]{\includegraphics[scale=0.19]{figs/vort4wn1ampt8000.png}}
	\subfigure[t=10500]{\includegraphics[scale=0.4]{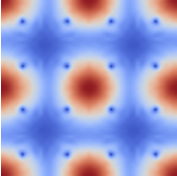}}
	\subfigure[t=21000]{\includegraphics[scale=0.4]{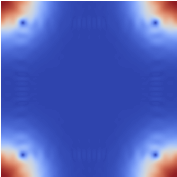}}
	\subfigure[t=31300]{\includegraphics[scale=0.4]{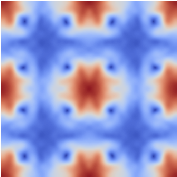}}
	\subfigure[t=41900=$T_P$]{\includegraphics[scale=0.4]{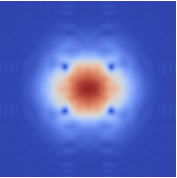}}\\
	\subfigure[$\omega_q$,t=0]{\includegraphics[scale=0.41]{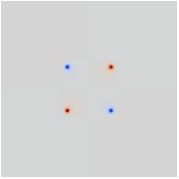}}
	\subfigure[t=8000]{\includegraphics[scale=0.195]{figs/vort4wn1enst8000.png}}
	\subfigure[t=10500]{\includegraphics[scale=0.4]{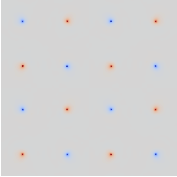}}
	\subfigure[t=21000]{\includegraphics[scale=0.4]{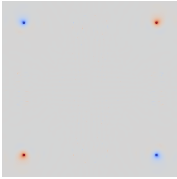}}
	\subfigure[t=31300]{\includegraphics[scale=0.4]{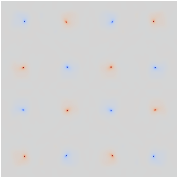}}
	\subfigure[t=41900=$T_P$]{\includegraphics[scale=0.4]{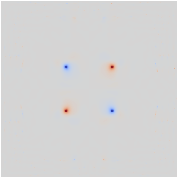}}\\
	\subfigure[ $\theta$,t=0]{\includegraphics[scale=0.41]{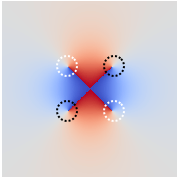}}
	\subfigure[ t=8000]{\includegraphics[scale=0.19]{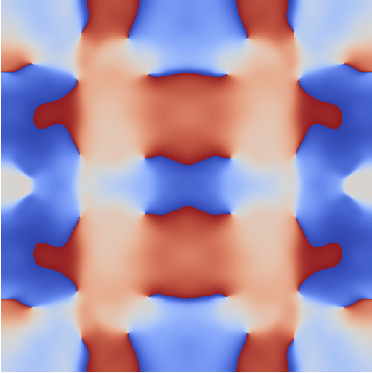}}
	\subfigure[ t=10500]{\includegraphics[scale=0.4]{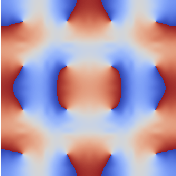}}
	\subfigure[ t=21000]{\includegraphics[scale=0.4]{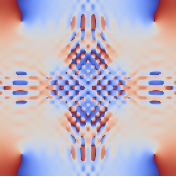}}
	\subfigure[ t=31300]{\includegraphics[scale=0.4]{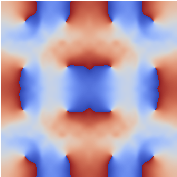}}
	\subfigure[ t=41900=$T_P$]{\includegraphics[scale=0.4]{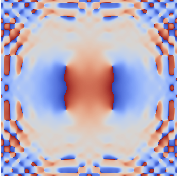}}\\	
	\caption{\label{fig:poinvortwn1}\footnotesize
	Poincar\'e recurrence for vortices with winding number $1$. The self-similar structure at $t=10500$ and $t=31300$ is most notable in the vorticity distribution plot, which is caused by the symmetric distribution of the vortices mimicking the initial condition. At $t=41900$, the amplitude $|\psi|$ and vorticity $\omega_q$ are nearly identical to the initial state. However, the phases of vortices undergo different shifts: $\pi/4$ for counter-clockwise rotating vortices and $-\pi/4$ for clockwise rotating vortices.  Spatial grid $512^2$.}
\end{figure*}

\begin{figure*}[!h!t!b!p]
	\subfigure[$|\psi_{tot}|$,t=0]{\includegraphics[scale=0.53]{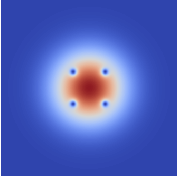}}
	\subfigure[t=10500]{\includegraphics[scale=0.65]{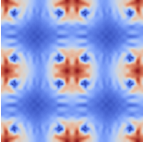}}
	\subfigure[t=21000]{\includegraphics[scale=0.65]{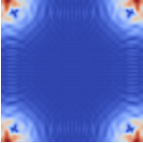}}
	\subfigure[t=31300]{\includegraphics[scale=0.65]{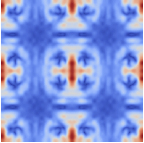}}
	\subfigure[t=41900]{\includegraphics[scale=0.65]{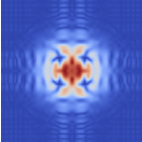}}\\
	\subfigure[$\omega_q$,t=0]{\includegraphics[scale=0.53]{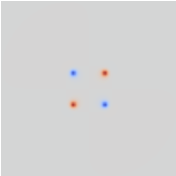}}
	\subfigure[t=10500]{\includegraphics[scale=0.65]{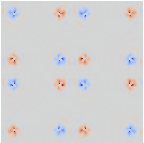}}
	\subfigure[t=21000]{\includegraphics[scale=0.65]{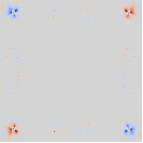}}
	\subfigure[=31300]{\includegraphics[scale=0.65]{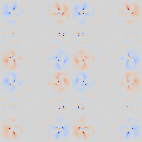}}
	\subfigure[t=41900]{\includegraphics[scale=0.65]{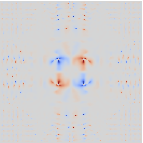}}\\
	
	\subfigure[ $\theta$,t=0]{\includegraphics[scale=0.53]{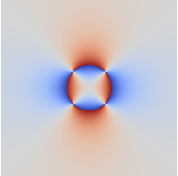}}
	\subfigure[ t=10500]{\includegraphics[scale=0.65]{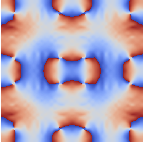}}
	\subfigure[ t=21000]{\includegraphics[scale=0.65]{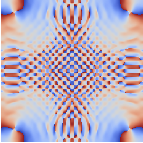}}
	\subfigure[ t=31300]{\includegraphics[scale=0.65]{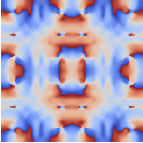}}
	\subfigure[  t=41900]{\includegraphics[scale=0.65]{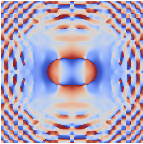}}\\	
	\caption{\label{fig:poinvortwn2}\footnotesize
	Poincar\'e recurrence for winding number $2$. The wave function evolves in a similar pattern as the $n=1$ case. At $t\sim10500$ and $t\sim31300$, a lattice of vortices is formed. At $t\sim21000$, the wave function restores to its initial distribution with a shifted origin. At $t=41900$, initial wave function is reproduced. However, more background noise is generated, such as the vortices with small depth indicated by the small dots in (h).  Spatial grid $512^2$}
\end{figure*}

For the first set of numerical simulations, we assume an initial wave function with vortices embedded in an inhomogeneous Gaussian BEC background. The total angular momentum is chosen to be zero and periodic boundary conditions are enforced. To satisfy these two constraints, the total wave function is taken as the product of the vortex wave functions: $\prod_i\psi_i(\mathbf{r})$, where the wave function of each vortex $\psi_i$ takes the form 
%
\begin{subequations}
\label{eq:gaussian}
\begin{eqnarray}
\displaystyle \psi_{tot}(\mathbf{r}, 0) &=&h\,e^{-a\,w_g\,r^2}\prod_{i=1}^4 \psi_i(\mathbf{r}-\mathbf{r}_i)\\  
\psi_i & =&\tanh(\sqrt{a}|\mathbf{r}-\mathbf{r}_i|)e^{\pm i\,n\,Arg(\mathbf{r}-\mathbf{r}_i)}.
\qquad
\end{eqnarray}
\end{subequations}
$|\psi_{i}(\mathbf{r}_i)|=0$ at the vortex core itself, $n$ is the winding number and we first consider $4$ vortices embedded in the Gaussian background. 
$h$ controls the wave function amplitude and $a$ can be viewed as a spatial rescaling parameter to resolve flow structures in the turbulence.
In Fig.~\ref{fig:poinvortwn1} and ~\ref{fig:poinvortwn2} we show the initial conditions for winding number $n=1$ and winding number $n=2$. The parameters are $h=0.05,\, a=0.01,\,w_g=0.01,\,g=5.0.$ The distance between neighboring vortices is: $L/4$ for $n=1$, and $2L/11$ for $n=2$ winding numbers. From the vorticity and phase plot, the location of the vortices are immediately pinpointed. The blue dots in enstrophy plots indicate clockwise rotating vortices [white circles in phase plot Fig.~\ref{fig:poinvortwn1} (m)], while red dots indicate counter-clockwise rotating vortices (black circles in the same phase plot). In essence the winding number $n=2$ vortex is a doubly degenerate simple vortex, with phase change $4 \pi$. Grid size is $512^2$.

The energy ratio parameter $\gamma=0.018$ for winding number $n=1$ vortices,  and $\gamma=0.0036$ for winding number $n=2$ vortices. The total energy is conserved to 7 significant digits in the simulation for 50000 iteration.  

\begin{figure}[!h!t!b!p]
	\subfigure[$n=1, E_I$]{\includegraphics[scale=0.65]{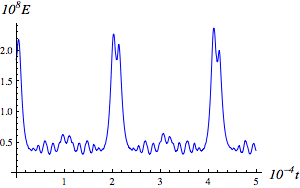}}
	\subfigure[$n=2, E_I$]{\includegraphics[scale=0.65]{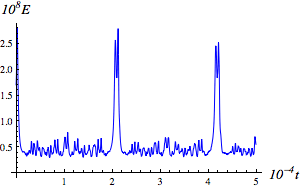}}
	\caption{\label{fig:vortinte}\footnotesize
	The time evolution of the internal energy for winding number (a) $n=1$ and (b) $n=2$ vortices. The first peak corresponds to the semi-Poincar\'e period $T_P/2$ and second peak corresponds to the full Poincar\'e period $T_P$, where $T_P = 41900$ for a spatial grid of $512^2$. The fluctuations in the internal energy are stronger for $n=2$ vortices, indicative of more vortex annihilations and creations for the higher winding number case.}
\end{figure}

\begin{figure}[!h!t!b!p]
	\subfigure[$|\psi_{tot}|$, t=8000]{\includegraphics[scale=0.2]{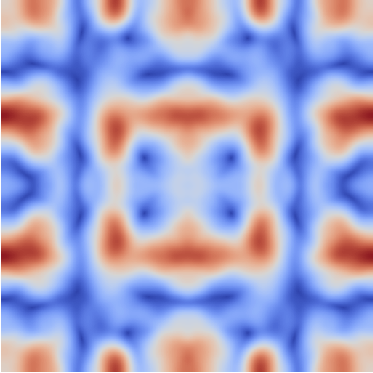}}
	\subfigure[vorticity $\omega_q$, t=8000]{\includegraphics[scale=0.2]{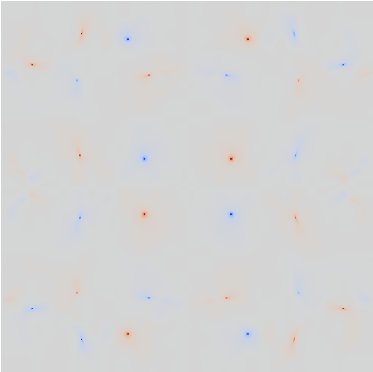}}
	\subfigure[phase $\theta$, t=8000]{\includegraphics[scale=0.2]{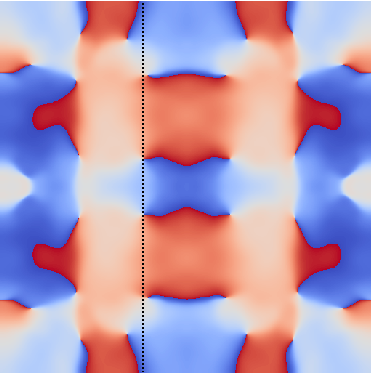}}\\
	\subfigure[ line plot near vortex core.]{\includegraphics[scale=0.6]{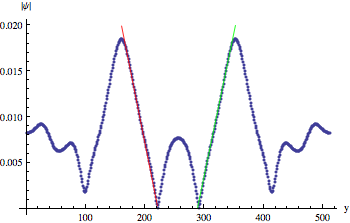}}
	\caption{\label{fig:vort4wn1t8000}\footnotesize
	The spatial distribution of the amplitude, vorticity and phase at t=8000. A clear symmetry is observed between quadrants which is due to the periodic boundary conditions and the symmetry in the initial conditions. However, within each quadrant, no vortex lattice is formed, which is most evident in the amplitude plot (a), the vorticity plot (b) and the phase plot (c).  Plot (d) is a 1D plot of $|\psi|$ along the $y$-axis at $x=197$, as indicated by the black dashed line in the phase plot (c). This line passes through a vortex pair (as identified by  $|\psi|=0$) . The solid lines in (d) show $|\psi| \sim r$ near the core region. The slopes are: $-3.44\times 10^{-4}$ for the red line and $3.46\times 10^{-4}$ for the green line, as expected from symmetry }
\end{figure}

For such low internal to kinetic energies, we find very short  Poincar\'e recurrence of the initial conditions.  This is very evident in the time evolution  of the internal energy, c.f. Fig.~\ref{fig:vortinte}, with the recurrence of the strong initial peak.
Based on the time evolution of internal energy, it is most instructive to examine the spatial distribution of the wave function amplitude $|\psi_{tot}|$, the density weighted vorticity $\omega_q$ and the phase $\theta$ at times $t=10500$, $t=21000$, $t=31300$ and $t=41900$, where $T_P = 41900$, Fig.~\ref{fig:poinvortwn1}. 
A regular lattice of vortices only occurs at integer multiples of the half-Poincare time,  $m T_P/2$.  For example at $t=8000$,  Fig.~\ref{fig:vort4wn1t8000},  no such self-similar structure is seen.  This is characteristic for all times away from integer $m T_P/2$.  Of course, there is an overall 4 quadrant symmetry because of the intial symmetry which is faithfully preserved by the QLA algorithm throughout the run.

At $t=10500$ and $t=31300$, 16 vortices are present in a highly symmetric lattice pattern, which is simply a four-fold version of the initial wave function. At $t=21000$, the distribution of $|\psi_{tot}|$ and $\omega_q$ would be same as the initial condiiton if the origin of the domain was shifted to $\mathbf{r}_0\rightarrow\mathbf{r}_0+L/2\mathbf{e}_x+L/2\mathbf{e}_y$ with $L$ being the domain size. At $t=41900$, the Poincar\'e recurrence period $T_P$, the distribution of both $|\psi|$ and $\omega_q$ closely approximates the initial state except for some sound wave interference particularly near the boundaries.   In the spatial distribution of the phase information $\theta$ there are phase shifts $\delta$ at the branch points/point vortices at $T_P$. In particular, one finds a phase shift of $\delta = \pi/4$ for counter-clockwise rotating vortices while having a phase shift $\delta = -\pi/4$ for clockwise rotating vortices, thus yielding different branch cuts connecting the same branch points.

Initial vortices with winding number $n=2$ are energetically unstable and will rapidly split into \textit{two} $n=1$ vortices \cite{pethick2002}. Therefore the Poincar\'e recurrence should now be viewed as the reproduction of that state immediately following the degenerate vortex splitting. To compare and contrast the evolution of the initial winding number $n=2$ vortices with the dynamics of the Poincar\'e recurrence for vortices of winding number $n=1$, we again consider the dynamics at $t=10500$, $21000$, $31300$, and $41900$. Fig.~\ref{fig:poinvortwn2} illustrates the global similarities and local differences with the case of winding number  $n=1$ vorticies, Fig.~\ref{fig:poinvortwn1} . Around $t\sim 10500$ and $t\sim 31300$, a vortex lattice again is populated by 16 pairs of vortices. At the semi-Poincar\'e period $T_P/2$, the wave function resembles the initial state but with the origin being shifted to $(L/2,L/2)$. It is assuming to point out that the Arnold cap map also exhibits this initial condition mirror symmetry at $T_P/2$ with full Poincare recurrence at $T_P$ - but only for certain initial conditions \href{http://en.wikipedia.org/wiki/File:Arnold%27s_Cat_Map_animation_(74px,_zoomed,_labelled).gif}{wiki}.  For other initial conditions one does not see this mirror symmetry at $T_P/2$. At the full Poincar\'e period $T_P$, one recovers the initial degenerate vortex split state but with considerably higher background noise than for the $n=1$ winding number case.  This increased noise level can be attributed to the higher density of vortices.

\begin{figure}[!h!t!b!p]
	\subfigure[$\theta$, t=10500]{\includegraphics[scale=0.35]{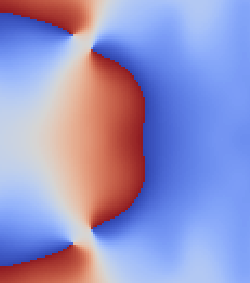}}
	\subfigure[$\theta$, t=10600]{\includegraphics[scale=0.35]{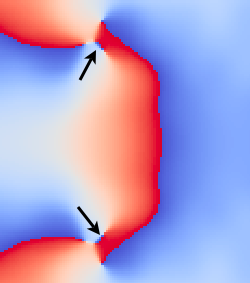}}
	\caption{\label{fig:qkh}\footnotesize
	Quantum Kelvin-Helmholtz instability in a localized region on a $512^2$ grid. The plots depicts the phase distribution at t=10500 and 10600 for initial winding number  $n=2$ vortices. The zoomed-in spatial domain is $[-256,-128]\times[-128,128]$. At t=10600, a pair of counter-rotating vortices are generated between the neighboring vortices with the same rotation, which can be identified by the new branch cuts identified by the black arrows. }
\end{figure}

We make a very interesting observation around $t=10500$ for the initial winding number $n=2$ vortices: we find that a pair of counter-rotating vortices are generated between neighboring vortices with the same rotation, c.f. Fig.~\ref{fig:qkh}.  Now in the spatial region between neighboring vortices with the same sense of rotation, the flow undergoes strong shear.  When this shear becomes sufficiently strong and exceeds a critical value new vortices are spun off by the  Quantum Kelvin-Helmholtz (QKH) instability \cite{takeuchi10}.  Our simulation confirms the recent postulate \cite{henn09, helsinki} that the quantum Kelvin-Helmholtz instability can be one important mechanism for quantum vortex generation. 
\subsection{Random phase initial condition}\label{subsec:rndinit}
\begin{figure}[!h!t!b!p]
	\includegraphics[scale=0.55]{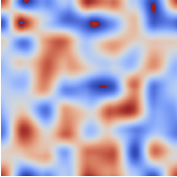}
	\caption{\label{fig:randinit}\footnotesize
	Bicubic fitted initial random phase $\theta(\mathbf{r},0)$ on a $512\times512$ lattice, with $m=8$.  The range of the phase is $[-\pi,\pi]$, and the 1st order derivative of the interpolate is constructed to be continuous.}
\end{figure}

We now consider an initial wave function with uniform density $\rho$ but with random phase $\theta$, $\psi = \sqrt{\rho} e^{i�\theta}$. The evolution of such  a wave function under the GP equation is energetically unstable. Randomly distributed vortices will be generated rapidly to form turbulence. To generate the random phase throughout the lattice domain, a bicubic interpolation \cite{keys1981} algorithm is employed. In this interpolation the desired 2D function $f(x,y)$ is approximated by the polynomial $p(x,y)$ defined on a (normalized) unit square:
\begin{equation}\label{eq:bicubicpoly}
p(x,y)=\sum_{i=0}^3\sum_{j=0}^3 a_{i,j}x^iy^j.
\end{equation}
The 16 unknown coefficients $a_{i,j}$ are determined by enforcing continuity at the 4 corners of the unit square. Typically, the continuity conditions are chosen to ensure that $f(x,y)$, $\partial_x f(x,y)$, $\partial_y f(x,y)$, $\partial_{x,y} f(x,y)$ be continuous at these 4 corners. This yields the required 16 equations to determine the coefficients $a_{i,j}$ uniquely. Thus, 
\begin{itemize}\setlength{\itemsep}{-0.3cm}
\item Discretize the domain into $m\times m$ unit squares. $m$ is known as the fragmentation level, with greater phase fluctuations for higher $m$;\\
\item Generate 4 pseudo-random numbers at each corner of the unit square, giving 16 random numbers on the unit square;\\
\item Compute the coefficients $a_{i,j}$ with the given 16 random numbers;\\
\item Periodicity is enforced by equating $a_{i,j}$ at the boundaries of the domain.
\end{itemize}
A random phase initial condition, shown in Fig.~\ref{fig:randinit}, is thereby constructed with $m=8$ on a $512\times512$ domain.

\begin{figure}[!h!t!b!p]
	\subfigure[]{\includegraphics[scale=0.65]{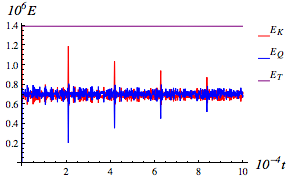}}
	\subfigure[]{\includegraphics[scale=0.65]{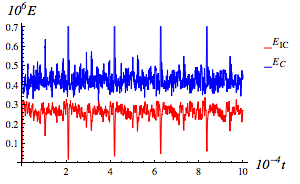}}
	\caption{\label{fig:randeng}\footnotesize
	Random phase initial conditions.  (a): Evolution of the total energy $E_T$, the kinetic energy $E_K$ and the quantum energy $E_Q$. (b): Evolution of the incompressible kinetic energy $E_{IC}$ and compressible kinetic energy $E_C$. The spikes in the kinetic energy are unmistakably associated with the occurrence of short Poincar\'e recurrence.  This signature is also observed in (b) with the sharp dips in the incompressible energy $E_{IC}$.}
\end{figure}

\begin{figure}[!h!t!b!p]
	\subfigure[phase $\theta(\mathbf{r})$ at t=0]{\includegraphics[scale=0.42]{figs/randphsini.png}}
	\subfigure[t=10500]{\includegraphics[scale=0.42]{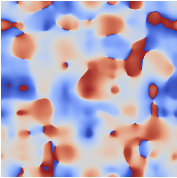}}\\
	\subfigure[t=21000]{\includegraphics[scale=0.42]{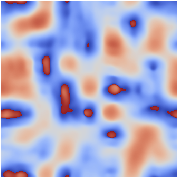}}
	\subfigure[t=41900]{\includegraphics[scale=0.42]{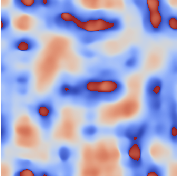}}\\
	\caption{\label{fig:poinrnd}\footnotesize
	Poincar\'e recurrence with random phase as initial condition. Unlike our earlier case with initial quantum vortices embedded in a Gaussian cloud, no vortex lattice formation is observed at t=10500 (see Fig.~\ref{fig:poinvortwn1}). The vortices are randomly distributed and resemble nothing like the initial (vortex-free) state. At t=21000, the vortices are depleted from the system which now bears some similarity with the initial condition. At t=41900, the vortices disappear again and the global features approximate the initial state.}
\end{figure}

To probe the occurrence of short Poincar\'e recurrence, we perform a long-time integration of our QLA representation of GP equation to $t=100000$ such that the ratio of internal energy to kinetic energy is $\gamma=0.00287$.. In  Fig.~\ref{fig:randeng} we show the evolution of the mean kinetic $E_K$ and quantum $E_Q$ energies. (The total energy $E_T$ is conserved to 10 significant digits throughout the simulation.) 
In the time evolution of the kinetic energy $E_K$ one sees strong spikes at t=21000, t=41900, t=63080, t=83980 ... with the spike amplitude decreasing. The corresponding phase plots at t=21000 and t=41900 are shown in Fig.~\ref{fig:poinrnd}. 

\begin{figure}[!h!t!b!p]
	\subfigure[vorticity $|\omega_q|$, t=20600]{\includegraphics[scale=0.75]{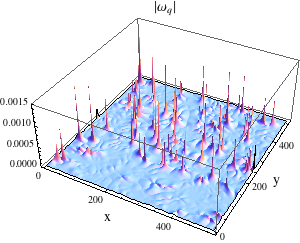}}
	\subfigure[vorticity $|\omega_q|$, t=20800]{\includegraphics[scale=0.75]{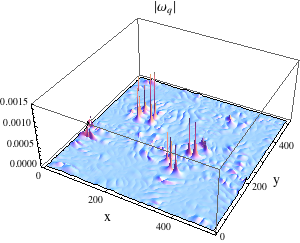}}
	\subfigure[vorticity $|\omega_q|$, t=21000]{\includegraphics[scale=0.75]{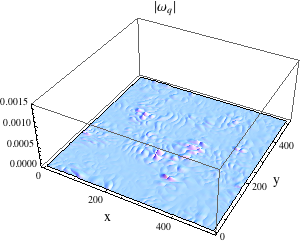}}
	\caption{\label{fig:randiniens}\footnotesize
	The distribution of the (both co- and counter- rotating) vortices at times $t = 20600$ and $t=20800$ before their total (but momentary) disappearance at $t=21000$. At most times in our simulation, the distribution of vortices resemble that at $t=20600$: a large number of vortices randomly distributed throughout the whole domain. However, the number of vortices rapidly tends to zero as $t$ approaches the semi-Poincar\'e period $21000$. }
\end{figure}

At t=10500, which corresponds to $T_P/4$ in Sec.~\ref{sec:vortinit}, no vortex lattice is observed. This can be explained by the fact that the vortices are not symmetrically distributed initially (in our current case there were no initial vortices). Somewhat unexpectedly, the vortices in the system are totally absent at t=21000, the semi-Poincar\'e period $T_P/2$. A dramatic decrease in the incompressible kinetic energy $E_{IC}$ is also observed, c.f. Fig.~\ref{fig:randeng}(b).  At t=41900, the Poincar\'e period $T_P$, the phase distribution is globally restored to its initial state (other than a global shift). Again, the vortices disappear from the system with strong dips in the incompressible kinetic energy $E_{IC}$.  The decrease in the $E_K$ spikes can be attributes to the loss of fine local structure in the phase $\theta(\mathbf{r})$. At other instances, we observe random distributed vortices upon which the quantum turbulence is developed. Fig.~\ref{fig:randiniens} illustrates the vortices distribution between $t=20600$ and $t=20800$, right before their total disappearance at $t=21000$. At $t=20600$, a large number of randomly distributed vortices can be clearly detected which characterizes the 2D quantum turbulence. At $t=20800$ the number of vortices is sharply reduced.  
%

\subsection{Loss of Short Poincar\'e recurrence}
\begin{figure*}[!h!t!b!p]
	\subfigure[Evolution of $E_T$, $E_K$, $E_Q$; $\gamma=0.0567$]{\includegraphics[scale=0.6]{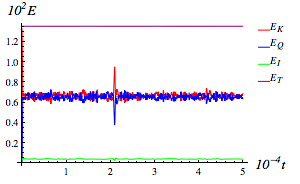}}
	\subfigure[Evolution of $E_T$, $E_K$, $E_Q$; $\gamma=0.133$]{\includegraphics[scale=0.6]{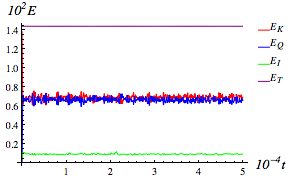}}\\
	\subfigure[Evolution of Enstrophy $Z$; $\gamma=0.0567$]{\includegraphics[scale=0.6]{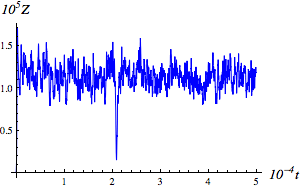}}
	\subfigure[Evolution of Enstrophy $Z$; $\gamma=0.133$]{\includegraphics[scale=0.6]{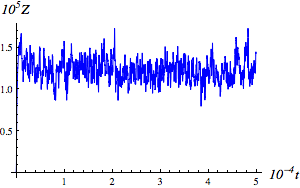}}
	\caption{\label{fig:poindes}\footnotesize
	Loss of Poincar\'e recurrence as $\gamma$ increases, random phase initial condition. The grid size is $512\times 512$ for this simulation, maximum iteration time is 50,000. The sharp drop of enstrophy in Fig(c) indicates the depletion of vortices from the system, which happens around $t\sim 21000$, the semi-Poincar\'e period.}
\end{figure*}

Short Poincar\'e recurrence times occur when the ratio of internal energy to kinetic energy is very small: $\gamma\sim\mathcal{O}[10^{-2}]$. However, as this ratio increases, the effects of the short Poincar\'e recurrence are weakened. With the random phase initial condition, our simulations show that for $\gamma\sim\mathcal{O}[10^{-1}]$, the short Poincar\'e recurrence can no longer be observed.  For $ t > 0 $ , many randomly distributed quantum vortices are formed.  As can be seen from Fig.~\ref{fig:poindes}, when $\gamma=0.0567$, there is a strong depletion of vortices  at $T_P/2$ as the state returns towards its initial state of no vortices (see also the strong dip in the enstrophy). However, there are no such signatures in the energies and enstrophy at $T_P$. As $\gamma$ increases to $0.133$, no depletion of vortices can be observed, which signals the loss of short Poincar\'e recurrence.  




\section{Energy spectra of 2D quantum turbulence}\label{sect:spec}

\begin{figure*}[!h!t!b!p]
	\subfigure[winding number $n=1,\, s_{ic}(t)$]{\includegraphics[scale=0.65]{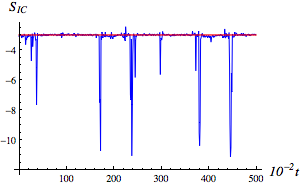}}
	\subfigure[winding number $n=2,\, s_{ic}(t)$]{\includegraphics[scale=0.65]{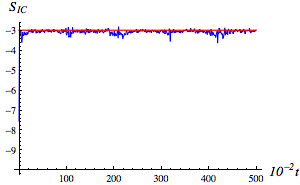}}
	\caption{\label{fig:vort4slope}\footnotesize
	The slopes of incompressible energy spectra. The fitting range for $s_{ic}$ is $k\in[50,100]$. The red line indicates $s_{ic}=-3$. The time averaged slopes are: $s_{ic}$=-3.23 for $n=1$ and $s_{ic}=-3.09$ for $n=2$. The variation of $s_{ic}(t)$ for $n=1$ is characterized by irregular decrease of $s_{ic}$, such as the slope near $t=24500$.}
\end{figure*}

\begin{figure*}[!h!t!b!p]
	\subfigure[t=24400, $s_{ic}=-3.005$]{\includegraphics[scale=0.55]{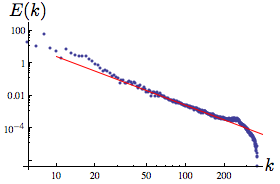}}
	\subfigure[t=24500, $s_{ic}=-5.828$]{\includegraphics[scale=0.55]{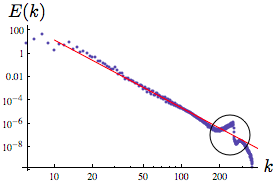}}
	\subfigure[t=24600, $s_{ic}=-3.357$]{\includegraphics[scale=0.55]{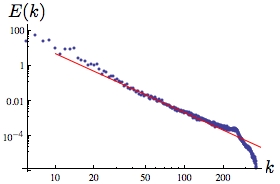}}
	\caption{\label{fig:vortwn1spec}\footnotesize
	Irregular incompressible kinetic energy spectrum for $n=1$. Fitting wave number range: $k\in[50,100]$. The spectral exponents are: (a) $s_{ic}=-3.005$; (b) $s_{ic}=-5.828$; (c) $s_{ic}=-3.357$. The kink (black circle) in (b) suggests sudden changes in the incompressible energy at small spatial scales.}
\end{figure*}

\begin{figure*}[!h!t!b!p]
	\subfigure[t=24400]{\includegraphics[scale=0.3]{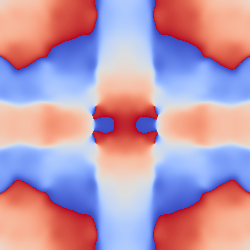}}
	\subfigure[t=24500]{\includegraphics[scale=0.3]{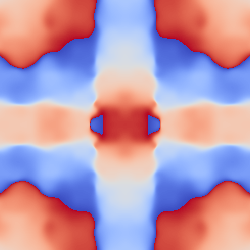}}
	\subfigure[t=24600]{\includegraphics[scale=0.3]{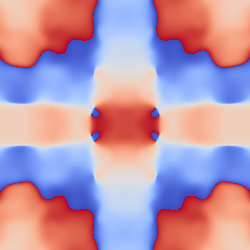}}
	\subfigure[centerleft blowup, t = 24400]{\includegraphics[scale=0.6]{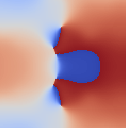}}
	\subfigure[centerleft blowup, t = 24500]{\includegraphics[scale=0.6]{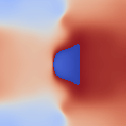}}
	\subfigure[centerleft blowup, t = 24600]{\includegraphics[scale=0.6]{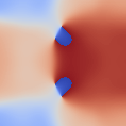}}
	\caption{\label{fig:vortwn1slopephs}\footnotesize
	Phase evolution $\theta(\mathbf{r},t)$ for $24400 < t <  24600$. The lower panel zooms into a centerleft part of the domain. At t=24500, (e), no branch cut can be observed (indicating no vortices present).  At t=24400, (d), 6 vortices can be identified, while at t = 24600, (f), there are 4 vortices.}
\end{figure*}

To study the energy spectra, Eq.(\ref{eq:spectra}),  in 2D quantum turbulence, we first consider the vortex initial conditions of Sec.~\ref{sec:vortinit}. For a 50000-iteration run on a $512^2$ grid, the spectra of incompressible and compressible kinetic energy are sampled every 100 iterations. If the energy spectra $\varepsilon(k)$ in Eq.\ref{eq:spectra} follow a power law, $\varepsilon(k)\propto k^{-\alpha}$, then the exponent $\alpha$ is immediately retrieved from $\partial_{\log(k)}\log[\varepsilon]$. Fig.~\ref{fig:vort4slope} displays the time variation of the incompressible spectral exponent $s_{ic}$ for two sets of initial conditions:  (a) winding number $n=1$ vortices, and (b) winding number $n=2$ vortices.  The red horizontal line corresponds to a $k^{-3}$ spectrum.
To understand the frequent appearance of very large spectral exponents for vortices with winding number $n=1$, we consider the spectrum around $t\sim24500$. From Fig.~\ref{fig:vortwn1spec}, it can be readily seen that the incompressible energy spectrum undergoes rapid change at very small spatial scales (i.e., at large $k$). 
This irregularity is also demonstrated by the phase plots $\theta(\mathbf{r})$ at $t=24400, 24500$ and $24600$ in Fig.~\ref{fig:vortwn1slopephs}.  One can readily see that it is the disappearance of vortices which is a major cause for the discontinuities in the incompressible energy spectrum.

However, with winding number $n=2$ vortices, the irregular variations in $s_{ic}$ are abated, c.f. Fig.~\ref{fig:vort4slope}. This could be explained by the presence of larger number of vortices in the system. With the number of vortices doubled, it is easier to reach a local critical velocity of counter flows to generate new pair of vortices; i.e., QKH occurs more frequently. Hence the probability of all the vortices to be annihilated from the system is extremely low. This is not dissimilar to what has been observed in 3D quantum turbulence \cite{vahala2be}. 

Following \cite{tsubota2010}, we now examine the effect of random phase initial condition to probe the cascades of 2D quantum turbulence by performing simulations on  $32768^2$ spatial grids with initial fragmentation level into blocks of $256 \times 256$. The total energy is conserved to 6 significant digits for 15000 iterations. The parameters in this simulation are:
\begin{align*}
	\text{coupling constant: }& g=1.0;\\
	\text{initial amplitude: }& |\psi(t=0)|=1.0;\\
	\text{coherence length: }& \xi=\frac{1}{\sqrt{g |\psi(t=0)|^2}}=1.0;\\
	\text{spatial resolution: }& \Delta x=0.03;\\
	\text{temporal resolution: }& a=\Delta x^2=0.0009;\\
	\text{size of system: }& L_s=32768\times\Delta x=983.04\xi.
\end{align*}
Here we identifying coherence length $\xi$ as is usually done in the literature, even though it is strictly defined only for small perturbations of an isolated vortex \cite{pethick2002} for a specific boundary value problem.

\begin{figure*}[!h!t!b!p]
	\subfigure[Evolution of $E_K$, $E_Q$ and $E_I$]{\includegraphics[scale=0.75]{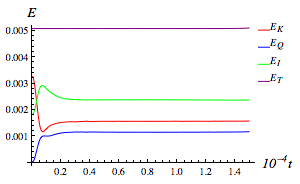}}
	\subfigure[$E_I$ v.s. $E_C$]{\includegraphics[scale=0.75]{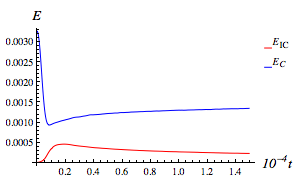}}
	\caption{\label{fig:mpeng}\footnotesize
	Evolution of energies for random initial condition.(a): Evolution of kinetic energy $E_K$, quantum energy $E_Q$, internal energy $E_I$ and total energy $E_T$. (b): Exchange between incompressible kinetic energy (red line) and compressible kinetic energy (blue line).  Grid $32768^2$}
\end{figure*}

For a $15000$- iteration run, the ratio between internal energy and kinetic energy is $\gamma=1.55$. The time evolution of the total ($E_T$), kinetic ($E_K$), quantum ($E_Q$) and internal ($E_I$) energy is shown in Fig.~\ref{fig:mpeng}.  These time evolutions are not dissimilar to that presented by Ref.~\cite{tsubota2010} with their set of random initial phases. 
The dynamics can be categorized into 2 broad stages besides the usual vortex-vortex interactions of quantum turbulence:
%

$\bullet$ {\it Rapid generation of vortices.} Since the initial wave function with random phase  is very unstable (because of strong velocity field variations arising from $\bm{v} \sim \nabla \theta$), a large number of vortices are rapidly generated. This generation of vortices causes the originally homogeneous amplitude $\sqrt \rho$  to fluctuate, which results in rapid increase in the internal energy and quantum energy. Concurrently, the incompressible energy increases rapidly. Provided that the enstrophy also increases rapidly at this stage, we conjecture that the increase of incompressible energy is mainly due to the generation of vortices in the system.
 \\

$\bullet$ {\it Depletion of vortices.} In this stage the major energy exchange is the energy transfer from vortices to elementary excitations, such as long range sound waves. This can be confirmed from the decrease of the incompressible kinetic energy with a corresponding increase in the compressible energy with the total kinetic energy remaining constant, c.f. Fig \ref{fig:mpeng}.

\begin{figure}
	\includegraphics[scale=0.75]{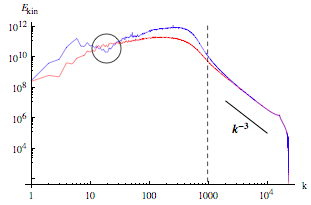}
	\caption{\label{fig:mprndspec}\footnotesize
	Spectra for incompressible (red) and compressible (blue) kinetic energy at $t=8000$. The unit of momentum $k$ is $\displaystyle k_u=\frac{2\pi}{L_s}$, with $L_s=983.04\xi$. The black dashed line indicates the position of $\displaystyle k_{\xi}=\frac{2\pi}{\xi}$.  The circle emphasizes the dip in the compressible energy which seems to propagate with time to smaller $k$ like a backward propagating pulse.  Grid $32768^2$}
\end{figure}

\begin{figure}
	\subfigure[Regions I and II]{\includegraphics[scale=0.8]{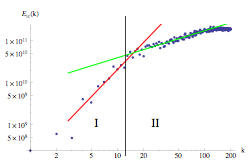}}\\
	\subfigure[Regions II and III]{\includegraphics[scale=0.8]{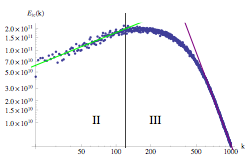}}\\
	\subfigure[Regions III and IV]{\includegraphics[scale=0.8]{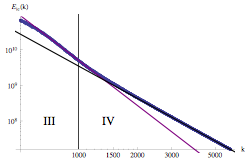}}
	\caption{\label{fig:mprndspecfit}\footnotesize
 Regression fits to the incompressible kinetic energy spectrum $k^{+\alpha}$ for various wave number windows.  {\bf(a)} Region $I$: $\alpha = +2.34$ for $1<k<15$ {\emph (red)}, Region $II$:$ \alpha =+0.65$ for $20<k<100$ {\emph (green)}; {\bf (b)} Region $II$: $\alpha = +0.65$ for $50<k<100$ {\emph (green)}, Region $III$:$ \alpha =-4.17$ for $600<k<1000$ {\emph (red)} ; {\bf (c)} Region $III$: $\alpha = -4.23$ for $700<k<1200$ {\emph (red)}, Region $IV$:$ \alpha =-3.03$ for $4000<k<5000$ {\emph (green)}.  Grid $32768^2$.} 
\end{figure}

At iteration step $t>3000$ (i.e., $0.3$ on the figure axis with units of $10^{-4}$), the only major energy exchange is between incompressible and compressible energy. We focus on iteration steps $t\in[4000,8000]$ to examine the spectra. Fig.~\ref{fig:mprndspec} depicts the energy spectra at $t=8000$ for both compressible and incompressible energy. The incompressible energy spectra can be broadly categorized into four regions: (I) $k \lesssim 0.01k_{\xi}$; (II) $0.01k_{\xi} \lesssim k \lesssim 0.1k_{\xi}$; (III) $0.1k_{\xi} \lesssim k \lesssim k_{\xi}$; (IV) $ k_{\xi} \lesssim k.$ The separation between (I) and (II) is signaled by the sudden drop of compressible energy at $k\sim15$ (the small dip encircled in Fig.~\ref{fig:mprndspec}). The regression fit of the incompressible energy spectrum is illustrated in Fig.~\ref{fig:mprndspecfit}. The variation of energy spectra is continuous in different regions and no bottle-neck effect are observed. It is interesting to notice that around $k\sim k_{\xi}$, the incompressible energy slope $\alpha$ is close to $-4.0$ and persists for a time interval $3000 < t < 14000$ during which the randomly distributed vortices dissipate away. 

To further examine the incompressible energy spectra near $k\sim k_{\xi}$, we sample the incompressible energy spectra every 50 iteration steps between $6000 < t < 10000$ within the wave number window $k\in[800,1200]$. The time averaged slope $\left<\alpha \right>=-4.145\pm0.066$. This is in good agreement with Saffman's $k^{-4}$ power law observed in Horng et. al. \cite{horng2009}. However, we do not observe any inverse energy cascade for $k<k_{\xi}$. 

\section{Conclusion}\label{sect:conclu}
We have investigated 2D quantum turbulence using a novel unitary QLA algorithm. The unitarity of the quantum algorithm makes it an efficient method for simulating the dynamics of GP system which demands the conservation of number of particles and total energy. The local collision and streaming operation enables the QLA to be parallelized almost ideally. The superlinear parallelization of the QLA allows us to probe the energy cascades with large grid simulation, such as $32768^2$. 

The spectra analysis of the incompressible kinetic energy reveals a ubiquitous $k^{-3}$ power law for large $k>k_{\xi}$. This power law, in 2D QT, may possibly be attributed to the result of Fourier Transform of the topological singularities  (and so may possibly be used to identify vortices in the 2D GP system).  However, there is still substantial sound waves at these very small scales due to the presence of a \emph{compressible} kinetic energy spectrum which overlays the incompressible spectrum for $k > k_{\xi}$.  At $k\sim k_{\xi}$, a $k^{-4}$ power law is observed, not dissimilar to that found by Horng et. al. \cite{horng2009} who then connect this to the Saffman $k^{-4}$ spectrum in CT.
However, no inverse energy cascade is observed in our simulation. This may attribute to the compressibility of GP system and fluctuation of enstrophy. During the simulations, we discover an unexpected short Poincar\'e recurrence provided that the ratio of internal energy to kinetic energy $\gamma\lesssim\mathcal{O}[10^{-2}]$. When $\gamma$ increases to the order of $\mathcal{O}[10^{-1}]$, this short Poincar\'e recurrence is no longer observed.

\begin{acknowledgments}
The authors would like to thank Jeffrey Yepez for many important and illuminating discussions on the quantum lattice gas algorithm.  This work was partially supported by the Department of Energy and the Air Force Office of Scientific Research.  The computations were performed on a William \& Mary cluster as well as at NERSC (DoE) and ERDC (DoD).
\end{acknowledgments}

\bibliography{qt2daps}

\end{document}